\documentstyle[aps,epsf,bbold]{revtex}
\newcommand{\bra}[1]{\mbox{$\langle #1|$}}
\newcommand{\ket}[1]{\mbox{$|#1\rangle$}}

\newcommand {\beq}
            {\begin{equation}}
\newcommand {\beqa}
            {\begin{eqnarray}}
\newcommand {\eeq}
            {\end{equation}}
\newcommand {\eeqa}
            {\end{eqnarray}}
\begin{document}
\begin{title}
{
\hfill{\small {\bf MKPH-T-99-17}}\\
{ \bf Coherent pion electroproduction on the deuteron in the 
$\Delta(1232)$ resonance region}\footnote{Supported by the Deutsche 
Forschungsgemeinschaft (SFB 443)}
}
\end{title}

\author{Thomas Ebertsh\"auser
 and Hartmuth Arenh\"ovel}
\address{
Institut f\"{u}r Kernphysik, Johannes Gutenberg-Universit\"{a}t,
       D-55099 Mainz, Germany}
\maketitle

\begin{abstract}
Coherent pion electroproduction on the deuteron is studied in the 
$\Delta(1232)$ resonance region in the impulse approximation, i.e., 
neglecting pion rescattering and two-body effects. The 
elementary reaction on the nucleon is described in the framework
of an effective Lagrangian approach including the dominant 
$P_{33}(1232)$ resonance and the usual background terms of the 
Born contributions for $\pi^0$ production. 
We have studied the influence of these different contributions on the 
various structure functions which determine the unpolarized 
exclusive differential cross section in a variety of kinematic regions. 
\end{abstract}

\section{Introduction}
Electromagnetic meson production off nuclei has become a major topic in 
medium energy physics. It offers the possibility to study the elementary 
production process in a strong interacting medium thus allowing to 
investigate in detail, for example, possible modifications of the 
elementary production process due to 
the surrounding medium. Another, though complementary aspect is 
to study the production on the neutron, which as a free process is in 
most cases not accessible. One exception is the radiative $\pi^+$ 
capture on the proton. Thus it is not surprising that in recent years, 
photo- and electroproduction of $\pi$ and $\eta$ mesons on  
nuclei have been studied intensively (see for example recent work in 
\cite{LeW97,EfH97}) 
with particular emphasis on testing the elementary 
production amplitude of the neutron. This 
renewed interest has been triggered primarily by the significant 
improvement on the quality of experimental data obtained with the new 
generation of high-duty cycle electron accelerators like, e.g., ELSA in Bonn 
and MAMI in Mainz \cite{BlB97,HRK97,LiB97,Mer98}. 

Pion photo- and electroproduction on the deuteron is especially 
interesting with respect to the above mentioned two complementary aspects 
since first it allows one to study this reaction on a bound nucleon in the 
simplest nuclear environment so that one can take into account medium 
effects in a reliable manner, at least in the nonrelativistic domain. 
Secondly, it provides important information on this reaction on the 
neutron. In view of this latter aspect, the deuteron is often considered 
as a neutron target assuming that binding effects can be neglected to a 
large extent. Most of the theoretical work has concentrated on the 
photo reaction (see Refs.\ \cite{ReS91,BlB93,GaG94,WiA95,WiA96,KaT97} 
where also references to earlier work can be found) while very few studies 
of the corresponding electroproduction process exist. 
The latter has been considered briefly in \cite{BlB93} with 
respect to target asymmetries for two specific kinematic settings, while in
\cite{JeS89} the role of exchange contributions in the longitudinal 
form factor near threshold has been investigated. But a systematic 
study is still missing.

The aim of the present work is to initiate such a systematic study of the 
$d(e,e'\,\pi^0)d$ reaction  from threshold through the $\Delta(1232)$ 
resonance similar to what has been done already for the corresponding photo 
production process \cite{WiA95,WiA96}. As a first step, we will restrict 
ourselves in this work to the impulse approximation (IA), where the 
production takes place at one nucleon only, in order to study 
details of the elementary reaction amplitude and, furthermore, we are 
interested mainly in the $\Delta$ resonance region. Because close to 
threshold the elementary operator used in this work does not 
give a realistic description. With respect to the 
elementary production operator,
we follow essentially the treatment of \cite{WiA95,WiA96} supplementing
the transverse current by charge and longitudinal contributions and 
including electromagnetic form factors at the elementary photon vertices. 
Any pion rescattering effects or two body contributions to the general 
current operator are neglected completely. Their treatment will be deferred 
to a later study. 
   
The paper is organized as follows: In the next section we present the 
theoretical framework for the coherent electroproduction process on 
the deuteron with definition of the structure 
functions which determine the differential cross section. 
Furthermore, we will briefly discuss a set of 
nonrelativistic amplitudes which form a basis for the expansion of 
the general $T$-matrix with appropriate invariant functions as coefficients.
A derivation of this set is given in the Appendix. 
Then we discuss in Sect.\ III the impulse approximation which 
serves as a calculational basis for the evaluation. In this section 
we also outline the elementary operator used in this work. 
In Sect.\ IV we present and discuss the results for the structure 
functions and form factors in various kinematic regions. 
Finally, we give a brief summary and 
an outlook. 

\section{General formalism}\label{formalism}

In this section we will briefly outline the formalism of coherent 
electroproduction of neutral pions on deuterium which in the 
one-photon approximation can be described as the absorption of 
a virtual photon, i.e.\
\beq
\gamma^{*}(k)+d(p)\rightarrow d(p')+\pi^0(q)\,,
\eeq
where the momenta of initial virtual photon and deuteron are denoted 
by $k$ and $p$, and the final deuteron and pion momenta by $p'$ and 
$q$, respectively. We will consider this reaction in the photon-deuteron 
($\gamma^{*} d$) c.m.\ frame. There we choose the $z$-axis along the 
virtual photon momentum $(\vec e_z= 
\hat k =\vec k /k)$, the $y$-axis parallel to $\vec k \times \vec q$ and 
the $x$-axis such as to form a right handed system. Thus the outgoing $\pi$ 
meson is described by the spherical angles $\phi$ and $\theta$  
with $\cos\theta=\hat{q}\cdot\hat{k}$. 

In the one-photon approximation
the differential cross section for the production process using 
unpolarized electrons and an unpolarized target is given by
\beqa
{d \sigma \over dk _2 ^{lab} d \Omega _e ^{lab} d 
\Omega _{\pi^0}^{c.m.}}&=&c\lbrace\rho_{L}f_{L}+\rho_{T}f_{T}
+\rho_{LT}f_{LT}\cos\phi+\rho_{TT}f_{TT}\cos 2\phi\rbrace
\,.\label{diffcross}
\eeqa
Here $c$ is a kinematic factor 
\begin{equation}
c(k_1^{lab},k^{lab}_2) = {\alpha \over 6 \pi^2} 
                          {k^{lab}_2 \over k_1^{lab} k_{\nu}^4}\;,
\end{equation}
where $k^{lab}_{1/2}$ denote incoming and scattered electron momenta 
in the lab frame, $\alpha$ the fine structure constant, and 
$k_{\nu} ^2=k_0^2-\vec k^{\,2}$ the four momentum transfer 
squared $(k = k_1 - k_2)$.
\beqa
\rho_L = -\beta^2 k^2_{\nu} {\xi^2 \over 2 \eta}\,,\hspace{.9cm}& &\hspace{1cm}
\rho_{LT} = -\beta k^2_{\nu} {\xi \over \eta} \sqrt{{\xi + \eta \over 8}}\;,
\\
\rho_T = -{1 \over 2} k^2_{\nu} (1 + {\xi \over 2 \eta})\,,& &\hspace{1cm}
\rho_{TT} =  k^2_{\nu} {\xi \over 4 \eta}\;,
\eeqa
with
\begin{equation}
\beta = {|{\vec k}^{\,lab}| \over |{\vec k}^{\,c}|},\,\,\,\,\,
\xi = -{k^2_{\nu} \over ({\vec k}^{\,lab})^{\,2}},\,\,\,\,\,
\eta = {\rm tan}^2({\theta_e^{lab} \over 2})\;,
\end{equation}
where $\theta_e^{lab}$ denotes the electron scattering angle in the lab system 
and $\beta$ expresses the boost from the lab system to the frame in which the 
$T$-matrix is evaluated and ${\vec k}^{\,c}$ denotes the momentum 
transfer in this frame. 

The structure functions in (\ref{diffcross}) are given in terms 
of the reduced reaction matrix elements $t_{m'\lambda m}$ which are defined
by the general $T$-matrix element of the current operator in the 
c.m.\ frame
\beqa
T_{m'\mu m}(W_{\gamma^\ast d},\theta,\phi)
&=&-\sqrt{\frac{\alpha |\vec q\,|E^{d}_k E^{d}_q}
{4\pi\, m_{d}\,W_{\gamma^\ast d}}}\,
\langle 1m'\!\mid\!\tilde{J}_{\mu}(\vec k-\vec q)
\!\mid\!1m\rangle \nonumber \\
&=&e^{i(\mu+m)\phi}\,t_{m'\mu m}(W_{\gamma^\ast d},\theta)\,,\label{Tmatrix} 
\eeqa
where the initial and final deuteron spin projections are denoted by $m$ 
and $m'$, respectively, the photon polarization by $\mu$, and initial 
and final deuteron c.m.\ energies by $E^{d}_k$ and $E^{d}_q$, 
respectively, with $E^d_p= \sqrt{m_{d}^{2}+\vec p^{\,2}}$ and $m_d$ 
as deuteron mass. Furthermore, 
$W_{\gamma^\ast d}$ denotes the invariant mass of the $\gamma^\ast d$ system 
and $k$ and $q$ the photon and $\pi^0$ momenta, respectively, 
in the $\gamma ^\ast d$ c.m.\ system. 

In detail one has
\begin{eqnarray}
f_{L}&=&\sum_{m',m}^{ }\Re e (t_{m'0m}^{*}t_{m'0m})\,,\hspace{1cm}
f_{T}=2\sum_{m',m}^{ }\Re e (t_{m'1m}^{*}t_{m'1m})\,,\nonumber\\
f_{LT}&=&4\sum_{m',m}^{ }\Re e (t_{m'0m}^{*}t_{m'1m})\,,\hspace{5mm}
f_{TT}=2\sum_{m',m}^{ }\Re e (t_{m'-1m}^{*}t_{m'1m})\,.
\end{eqnarray}
The structure functions are functions of the invariant mass 
$W_{\gamma^\ast d}$, the squared three momentum transfer 
$(\vec k^{\,c.m.})^2$ and the angle $\theta$ between $\vec k$ and 
the momentum $\vec q$ of the outgoing pion, i.e.\
\beq
f_\alpha=f_\alpha(W_{\gamma^\ast d},(\vec k^{\,c.m.})^2,\theta)
\quad\mbox{for}\quad \alpha\in\{L,T,LT,TT\}\,.
\eeq
The inclusive cross section obtained by integrating over the pion angles
is determined by the longitudinal and transverse form factors $F_{L}$ and
$F_{T}$, respectively, which are defined by
\beq
F_{L/T}(W_{\gamma^\ast d},(\vec k^{\,c.m.})^2)=\int d\Omega_q\,
f_{L/T}(W_{\gamma^\ast d},(\vec k^{\,c.m.})^2,\theta)\,.
\eeq
 
In analogy to the CGLN-amplitudes of e.m.\ pion production on a nucleon, a
set of 13 basic covariant amplitudes $\Omega_\alpha$ has been derived in 
\cite{Are98} which allow a representation of the $T$-matrix as a 
linear superposition of these amplitudes with invariant functions 
$F_\alpha(s,t,u)$ as coefficients depending on the Mandelstam variables only
\beq
T=\sum_\alpha F_\alpha(s,t,u)\,\Omega_\alpha\,. \label{mficov}
\eeq
An equivalent nonrelativistic set of transverse ${\cal O}_{T,\beta}$
and longitudinal ${\cal O}_{L,\beta}$ operators has also been given in 
\cite{Are98} which is listed in Table \ref{tabkqls} explicitly. Here 
$\vec S$ denotes the spin operator for $(S=1)$-particle and $S^{[2]}$
the corresponding tensor operator defined by
\beq
S^{[2]}=[S^{[1]}\times S^{[1]}]^{[2]}
\eeq
in the notation of Fano and Racah \cite{FaR59}. The operators 
${\mathbb 1}_3$, $S^{[1]}$, and $S^{[2]}$ form a complete set of 
$(3\times 3)$-matrices in $(S=1)$-space. Furthermore, in Table \ref{tabkqls} 
$\hat k$ and $\hat q$ denote corresponding unit vectors and with 
$\hat k=k^{[1]}$
\beq
k^{[2]}=[k^{[1]}\times k^{[1]}]^{[2]}\,.
\eeq
In the Appendix, we give an independent proof that this set forms a 
complete basis for the representation of the $T$-matrix for the 
process under consideration. In the present 
nonrelativistic framework this set is more appropriate and one can 
represent the $T$-matrix correspondingly
\beq
T_{m'\mu m}=\delta_{|\mu| 1}\,
\sum_{\beta=1}^9 f_\beta^T\,({\cal O}_{T,\beta})_{m'\mu m}
+\delta_{\mu 0}\,\sum_{\beta=1}^{4} 
f_\beta^L\,({\cal O}_{L,\beta})_{m'\mu m}\,
\eeq
with appropriate functions $f_\beta^{L/T}$. 
Any contribution to the reaction can 
be expanded in terms of these amplitudes. In the present work, however, 
where we restrict ourselves to the pure impulse approximation (IA), we
will evaluate directly the various terms of the elementary production 
operator on the nucleon without separating explicitly the various operator 
contributions. 

\section{The impulse approximation}
\begin{figure}
\centerline{%
\epsfysize=6cm
\epsffile{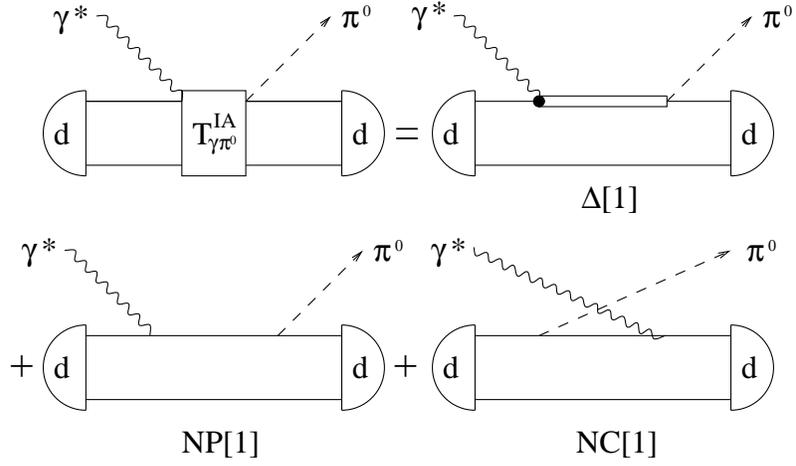}}
\vspace*{.5cm}
\caption{Impulse approximation diagrams for $\gamma^\ast + d \rightarrow 
\pi^0 + d$ with resonance contribution $\Delta[1]$, direct NP[1] and 
crossed NC[1] nucleon pole contribution.}
\label{figIA}
\end{figure}

In the IA, the $\pi^0$ production takes place at 
one nucleon while the other acts merely as a spectator. Thus the basic 
diagrams of the IA consisting of a resonance and two Born terms 
are the ones depicted in Fig.\ \ref{figIA}. 
Consequently, the production operator $\hat t_{\gamma\pi^0}^d $ for 
the reaction on the deuteron which governs the $T$-matrix in the c.m.\ 
frame of virtual photon and deuteron 
\beq
T_{m'\mu m}(W_{\gamma^\ast d},\theta,\phi)=\langle  -\vec q,m'|
\hat t_{\gamma^{\ast} \pi^0}^d(W_{\gamma^\ast d})| 
-\vec k,m \mu\rangle\,,\label{Tmatrixa}
\eeq
is obtained from the elementary operator 
$\hat t_{\gamma^{\ast}\pi^0 }$ by
\begin{equation}
\hat t_{\gamma^{\ast}\pi^0 }^{d}(W_{\gamma^\ast d})
=\hat t_{\gamma^{\ast}\pi^0}^{\,(1)}(W_{\gamma^\ast N}) 
\otimes {\mathbb 1}^{(2)} +{\mathbb 1}^{(1)} \otimes 
\hat t_{\gamma^{\ast}\pi^0 }^{\,(2)}(W_{\gamma^\ast N})\,,
\end{equation}
where the upper index in brackets refers to the nucleon on which the operator 
acts. Off-shell effects will be neglected. The invariant mass of the 
$\gamma^\ast N$ system is denoted by $W_{\gamma^\ast N}$. The assignment of 
$W_{\gamma^\ast N}$ will be discussed below.

Explicitly one finds
\beq
T_{m'\mu m}=2\sum_{m_s',m_s}
\int d^{3}p \,\psi^{\ast}_{m_s' m^{\prime}}
\Big(\vec{p}-\frac{\vec{q}}{2} \Big)\bra{\vec{p}-\vec{q}\,;1m_s',00}
\hat t^{(1)}_{\gamma^{\ast} \pi^0 }\ket{\vec{p}-
\vec{k}\,;1m_s,00}\psi_{m_s m}\Big(\vec{p}-\frac{\vec{k}}{2}  \Big),
\label{equ:t}
\eeq
where $\bra{\vec{p}^{\,\prime}\,;1m_s',00}
\hat t^{(1)}_{\gamma^{\ast} \pi^0 }\ket{\vec{p}\,;1m_s,00}$ denotes the 
elementary $t$-matrix for initial and final nucleon momentum  
$\vec{p}$ and $\vec{p}^{\,\prime}$, respectively, evaluated between
two-nucleon spin and isospin wave functions
$|1m_s',00\rangle=
|(\frac{1}{2}\frac{1}{2})1m_s',(\frac{1}{2}\frac{1}{2})00\rangle$ and 
$|1m_s,00\rangle$. Furthermore, $\psi_{m_s m}(\vec p\,)$
denotes the intrinsic part of the deuteron wave function 
in momentum space projected onto $|1m_s,00\rangle$ 
\beqa
\psi_{m_s m}(\vec p\,)&=&\sum_{l=0,2}\sum_{m_l}(l m_l 1 m_s|1 m) u_l(p)
Y_{l,{m_l}}(\hat p)
\,,
\eeqa
where $\vec p= \frac{1}{2}(\vec p_{1}-\vec p_{2})$ denotes the 
relative momentum of the two nucleons in the deuteron. 

For the evaluation of (\ref{equ:t}) one has to specify 
the elementary operator $\hat t_{\gamma^{\ast}\pi^0}$ 
for the reaction 
$\gamma^{\ast} + N \rightarrow \pi^0 + N$. As already mentioned, 
we will follow the treatment in \cite{WiA95,WiA96} for the coherent 
photoproduction process taking into account the $\Delta$ 
resonance contribution and as background the Born terms from the 
direct and crossed nucleon pole diagrams, denoted ``$NP$'' and 
``$NC$'', respectively, in Fig.\ \ref{figIA}, i.e.\
\beq
\hat t_{\gamma^{\ast}\pi^0}=\hat t^{(\Delta)}_{\gamma^{\ast}\pi^0}+
\hat t^{(NP)}_{\gamma^{\ast}\pi^0}+\hat t^{(NC)}_{\gamma^{\ast}\pi^0}\,.
\eeq
We will first consider the Born terms whose 
contributions are given by  
\beqa
\hat t^{(NP)}_{\gamma^{\ast}\pi^0}(z)&=&v^{\dagger}_{\pi^{0}N}\,
G^{(NP)}(z)\,v_{\gamma^{\ast}N}\,,\\
\hat t^{(NC)}_{\gamma^{\ast}\pi^0}(z)&=&v_{\gamma^{\ast}N}\,
G^{(NC)}(z)\,v^{\dagger}_{\pi^{0}N}\,.
\eeqa
The vertex functions are 
\beqa
v^{\dag}_{\pi^{0}N}&=&-i\frac{f_{\pi}}{m_{\pi}}\vec\sigma\!
\cdot\!\vec q\,\tau_{0}\,,\\
v_{\gamma^{\ast}N}&=&\epsilon_\nu(\mu) \,j^\nu_N(\vec k)\,,
\eeqa
where $\epsilon_\nu(\mu)$ denotes the polarization vector of the 
virtual photon with helicity $\mu$. As $\pi N$ coupling constant we 
have chosen $\frac{f_{\pi}^{2}}{4\pi}=0.08$, and the nucleon charge 
and current densities are given by the nonrelativistic expressions 
\beqa
\rho_{N}&=&{\hat e}\,,\\
\vec \jmath_{N}&=&\frac{\hat e}{2m_{N}}(\vec p\,'\!\!+\vec p\,)+
\frac{\hat e+\hat\kappa}{2m_{N}}i\vec\sigma\times\vec k\,,
\eeqa
with
\beq
\hat e=\frac{e}{2}(1+\tau_{0})\,F(k_{\mu}^2)\,,\qquad
\hat\kappa=\frac{e}{2}\left(\kappa_{p}(1+\tau_{0})
+\kappa_{n}(1-\tau_{0})\right)\,F(k_{\mu}^2)\,.
\eeq
Here $e$ denotes the elementary charge and $\kappa_{n/p}$ the 
anomalous magnetic moment of neutron and proton, respectively. 
Furthermore, we have introduced a common e.m.\ form factor 
$F(k_{\mu}^2)$ for which we haven chosen the dipole parametrization
\beq
F(k_{\mu}^2)=\left(1-\frac{k_{\mu}^2}{0.71(\text{GeV})^{2}}\right)^{-2}\,.
\eeq
The propagators of the direct and crossed terms are of the form
\beq
G^{(NP)}(z)=\frac{1}{z-E_{\vec p}^{(N)}+i\epsilon}\,,\qquad 
G^{(NC)}(z)=\frac{1}{z-E_{\vec p}^{(N)}-E^{(\pi)}-\omega+i\epsilon}\,,
\eeq
where $E^{(\pi)}=\sqrt{m_{\pi}^{2}+{\vec q}\,^{2}}$ denotes the 
relativistic pion energy and 
$E_{\vec p}^{(N)}\!=\!m_{N}+\frac{{\vec p}\,^{2}}{2m_{N}}$ the 
nonrelativistic nucleon energy. 

For the resonance contribution 
\beq
\hat t^{(\Delta)}_{\gamma^{\ast}\pi^0}(z)=v^{\dagger}_{\pi^{0}N\Delta}\,
G^{(\Delta)}(z)\,
v_{\gamma^{\ast}N\Delta}\,,
\eeq
we use for the $\gamma^\ast N\Delta$ 
vertex the dominant $M1$ contribution of the $N\Delta$ transition current 
\beqa
\rho_{\gamma^{\ast}N\Delta}^{M1}&=&eF(k_{\mu}^2)
\frac{\tilde G_{\gamma N\Delta}^{M1}(E_{\Delta})}{2m_{N}m_{\Delta}}
i(\vec\sigma_{\Delta N}\!\times\!\vec k)\!\cdot\!\vec p_{N}\,
\tau_{\Delta N,0}\,,\\
\vec \jmath_{\gamma^{\ast}N\Delta}^{\,\,M1} 
&=&eF(k_{\mu}^2)\frac{\tilde G_{\gamma N\Delta}^{M1}(E_{\Delta})}
{2m_{N}}i\vec\sigma_{\Delta N}\!\times\!\vec k_{\gamma N}\,
\tau_{\Delta N,0}\,,
\eeqa
neglecting the tiny $C2$ and $E2$ parts, because they are not expected to 
play a significant role in the pure impulse approximation where also other 
small effects are not considered. Here we use $m_{\Delta}=1232$ MeV and 
\beq
\vec k_{\gamma N}=\frac{m_{N}\vec k-\omega\vec p_{N}}{m_{\Delta}}\,,
\eeq
and $E_{\Delta}=W_{\gamma^\ast N}$ is the $\Delta$ energy in its 
rest system. The $\gamma^\ast N\Delta$ coupling is taken as energy 
dependent and parametrized in the form \cite{WiA95}
\beq
\tilde G_{\gamma N\Delta}^{M1}(E_{\Delta})=\left\{
\begin{array}{ll}
\mu_{M1}(q_{0}(E_\Delta))
\exp[i\Phi_{M1}(q_{0}(E_\Delta))] & \mbox{ if }\, 
E_{\Delta}>m_{\pi}+m_{N}\,,\\
\mu_{0} & \mbox{ else}\,,
\end{array}
\right.
\eeq
where the different quantities are defined as
\beqa
\mu_{M1}(q_{0})&=&\mu_{0}+\mu_{2}\left(\frac{q_{0}}{m_{\pi}}\right)^{2}
+\mu_{4}\left(\frac{q_{0}}{m_{\pi}}\right)^{4}\,,\\
\Phi_{M1}(q_{0})&=&\frac{q_{0}^{3}}{a_{1}+a_{2}q_{0}^{2}}\,,
\eeqa
with $q_0(E_\Delta)$ as the on-shell pion momentum in this 
system, given by 
\beq
E_{\Delta}=\sqrt{m_{\pi}^{2}+q_{0}^{2}}+m_{N}+\frac{q_{0}^{2}}{2m_{N}}\,. 
\eeq
The parameters have been fixed by fitting the $M^{3/2}_{1+}$ multipole
of pion photoproduction yielding the values 
$\mu_{0}=4.16$, $\mu_{2}=0.542$, $\mu_{4}=-0.0757$, $a_{1}=0.185$ fm$^{-3}$,
and $a_{2}=4.94$ fm$^{-1}$ \cite{WiA95}.

The $\pi^0 N\Delta$ vertex is given in the usual form 
\beq
v^{\dag}_{\pi^{0}N\Delta}=-i\frac{f_{\Delta}}{m_{\pi}}F_{\Delta}
(\vec q\,^{2}_{\pi N})\vec\sigma_{N\Delta}\!\cdot\!
\vec q_{\pi N}\,\tau_{N\Delta,0}\,,
\eeq
with
\beq
\vec q_{\pi N}=\frac{m_{N}\vec q-E^{(\pi)}\vec p_{N}}{m_{N}+E^{(\pi)}}\,,
\eeq
and the $N\Delta$ coupling constant $\frac{f_{\Delta}^{2}}{4\pi}=1.393$. 
Here $\vec\sigma_{N\Delta}$ and $\tau_{N\Delta,0}$ denote the usual spin 
and isospin $N\Delta$ transition operators, respectively. 
The hadronic form factor is taken of dipole type with parameters 
also obtained in the above mentioned fit
\beq
F_{\Delta}(\vec q\,^{2}_{\pi N})=
\frac{\Lambda_{\Delta}^{2}-m_{\pi}^{2}}
{\Lambda_{\Delta}^{2}+\vec q\,^{2}_{\pi N}}\,,
\eeq
where $\Lambda_{\Delta}=287.9$ MeV, 
and $m_{\Delta}^{0}=1315$ MeV. Finally, the $\Delta$ resonance 
propagator is of the form
\beq
G^{(\Delta)}(E_{\Delta})=\frac{1}{E_{\Delta}-M_{\Delta}(E_{\Delta})+
\frac{i}{2}\Gamma_{\Delta}(E_{\Delta})}\,,
\eeq
where the energy dependent mass and width are given by
\beqa
M_{\Delta}(E_{\Delta})&=&m_{\Delta}^{0}+\frac{f_{\Delta}^2}{12\pi^{2}
m_{\pi}^{2}}\,\wp\!\int_{0}^{\infty}\!\frac{dq'\,{q'}^{4}
F_{\Delta}^{2}({q'}^{2})}
{\sqrt{m_{\pi}^{2}+{q'}^{2}}(E_{\Delta}-\sqrt{m_{\pi}^{2}
+{q'}^{2}}-m_{N}-\frac{{q'}^{2}}{2m_{N}})}\,,
\eeqa
and
\beqa
\Gamma_{\Delta}(E_{\Delta})&=&\left\{
\begin{array}{ll}
\frac{q_{0}(E_{\Delta})^{3}m_{N}}{6\pi m_{\pi}^{2}
(\sqrt{m_{\pi}^{2}+q_{0}(E_{\Delta})^{2}}+m_{N})}
\,f_{\Delta}\,F_{\Delta}^{2}(q_{0}(E_{\Delta})^{2}) & \mbox{ if } 
E_{\Delta}>m_{\pi}+m_{N}\,,\\
0 & \mbox{ else}\,.
\end{array}
\right.
\eeqa

The elementary operator $\hat t_{\gamma^\ast \pi^0 }^{\,(1)}$ is a function 
of the photon, nucleon and $\pi^0$ momenta $\vec k$, $\vec p$, and 
$\vec q$, respectively, the photon polarization $\mu$, and of the 
invariant mass $W_{\gamma^\ast N}$ of the photon-nucleon subsystem. 
Implementing this operator into a bound system poses 
the problem of assigning an invariant mass 
$W_{\gamma^\ast N}$ for the struck or active nucleon. This question 
has been discussed in detail in \cite{BrA97} for coherent $\eta$ 
photoproduction on the deuteron by studying various prescriptions. 
In this work we have adopted the spectator-on-shell assignment as 
in \cite{WiA95,WiA96}.

\section{Results and Discussion}\label{kap:2}
\begin{figure}[h]
\centerline{%
\epsfysize=10cm
\epsffile{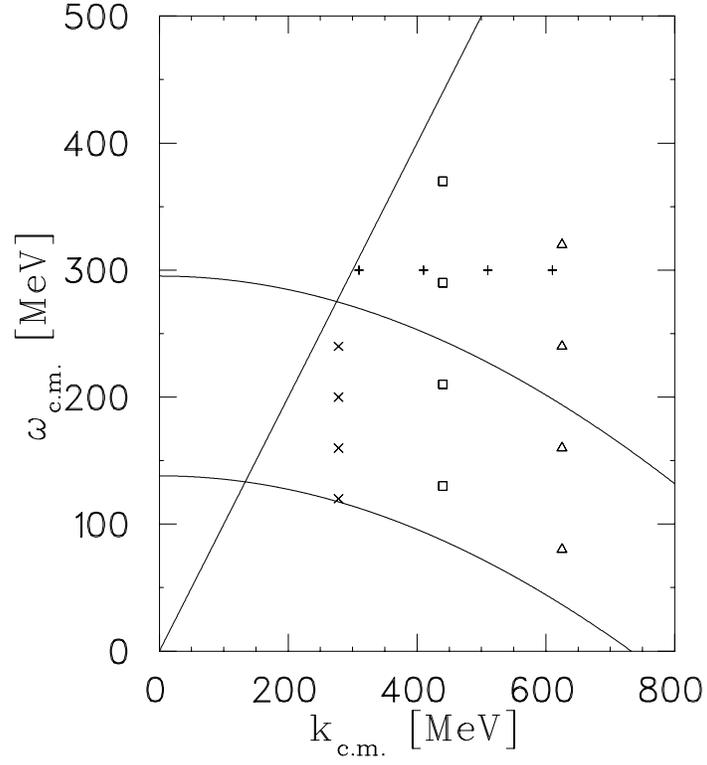}}
\vspace*{.5cm}
\caption{Various sets of kinematics in the $k$-$\omega$-plane chosen for the 
numerical evaluation, labeled $A\ (\times)$, $B\ (\Box)$, $C\ (\triangle)$ 
and $D\ (+)$. The straight line represents the photon line. The lower curve 
marks the $\pi^0$ production threshold while the upper curve represents 
the location of the $\Delta$-resonance.}
\label{k-omega-plane}
\end{figure}

Having fixed the parameters of the $\pi^0$ production model for 
the elementary reaction on the nucleon, 
we have calculated the coherent reaction on the deuteron. 
The $t$-matrix elements of (\ref{equ:t}) have been evaluated numerically 
using Gauss integration in momentum space. As deuteron wave function, we 
have taken the one of the Bonn r-space potential in the parametrized form 
\cite{MaH87}. In order to test the numerical program, we have first calculated 
the differential and total cross sections at the photon point and compared 
the results to the one of \cite{WiA95} for which we obtained complete 
agreement. Since in electroproduction energy and momentum transfers can 
be varied independently in the spacelike region, we have chosen various 
cuts along constant energy or momentum transfer as shown in Fig.\ 
\ref{k-omega-plane} for which we calculated the four structure functions. 
For the sets A, B and C we have fixed the momentum 
transfer $k$ and varied the energy transfer $\omega$, while for set D 
we have fixed $\omega$ and varied $k$. The set A at a moderate momentum 
transfer of $k^2=2\,\rm{fm}^{-2}$ covers the region right above the 
production threshold up to the onset of the $\Delta$-resonance. The sets B 
and C are cuts across the $\Delta$-resonance for fixed three momentum 
transfers $k^2=5\,\rm{fm}^{-2}$ and $k^2=10\,\rm{fm}^{-2}$, respectively, 
while set D for a fixed energy transfer $\omega=300$ MeV remains 
essentially above but close to the resonance region. 

\begin{figure}
\centerline{%
\epsfysize=18cm
\epsffile{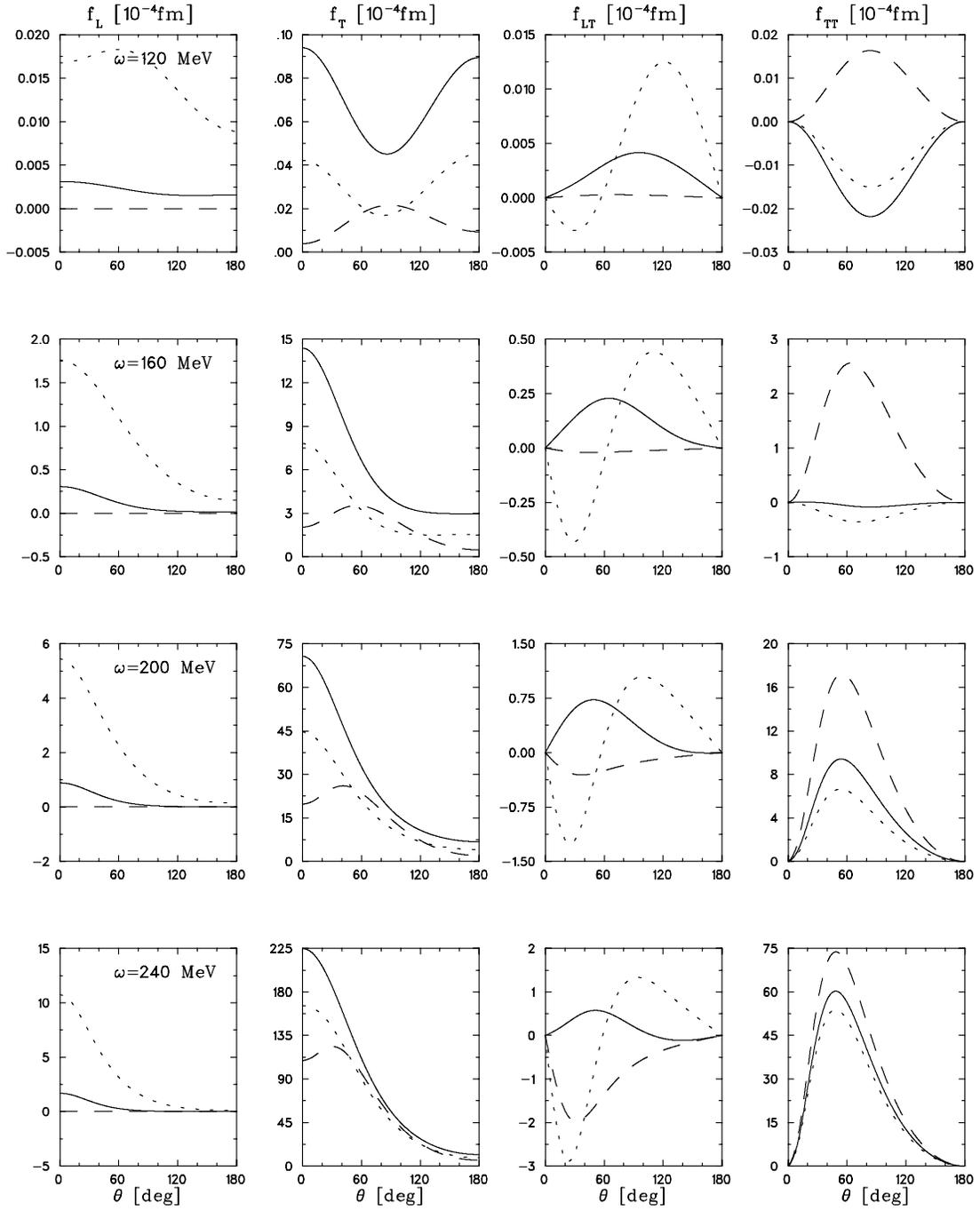}}
\vspace*{.5cm}
\caption{Structure functions for the set A at constant momentum transfer
$k^2=2\,\rm{fm}^{-2}$ for various energy transfers. 
Dashed curves: pure $\Delta$ contribution; 
dotted curves: direkt Born term added; full curves: complete IA. }
\label{2_fm-2}
\end{figure}

We will start the discussion of the structure functions with set A shown 
in Fig.\ \ref{2_fm-2}. The longitudinal structure function $f_L$ is almost
not affected by the $\Delta$ resonance. This is a consequence of the assumed 
pure $M1$ excitation of the $\Delta$ leading to a completely transverse 
current in the $\Delta$ rest frame, 
so that any longitudinal contribution which arises when going to other 
frames are negligible. Therefore, this structure function is governed by 
the Born terms only. One readily notes a strong destructive interference 
between direct and crossed terms resulting in a small forward peaked 
angular distribution, except very close to threshold. 

For the transverse structure function $f_T$ one notices an increasing 
forward peaking of the total IA 
with increasing $\omega$, except for the lowest energy 
transfer close to threshold where an almost symmetric forward backward 
peaking appears. At the lowest energy transfer, $f_T$ is dominated by the Born 
terms displaying a constructive interference between direct and crossed 
contributions in sharp contrast to the findings for $f_L$. This 
different behaviour of the Born terms 
is easily understood from the structure of 
the corresponding operators and the sign of the propagators. For the 
longitudinal part, the operators of direct and crossed contributions 
have the same sign, while they differ in sign for most of the 
transverse operators which combined with the opposite signs of 
the propagators leads to the observed destructive, respectively, 
constructive behaviour. Close to threshold, the 
$\Delta$ contribution is small as expected. It shows a slight peaking 
around 90 degree. With increasing $\omega$ this peak moves more and more 
to forward angles. Also its relative size increases rapidly becoming 
dominant at the highest energy displayed. At the higher energies $f_T$ 
is more than an order of magnitude bigger than $f_L$. 

For the longitudinal-transverse interference structure function $f_{LT}$ 
the $\Delta$ contribution yields a negative structure function of 
increasing size with increasing $\omega$. The direct Born term 
interferes constructively in the forward direction while destructively 
at backward angles and the crossover moves more and more to smaller 
angles with increasing energy. The crossed Born contribution interferes 
destructively with both, $\Delta$ and direct Born terms, 
resulting in a positive structure function over 
a large forward angular range. The size of $f_{LT}$ is comparable 
to $f_L$ for the total contributions.  

Finally, the transverse interference structure function $f_{TT}$ receives 
a positive contribution from the $\Delta$ with a peak moving from about 
90 degree near threshold to smaller angles, about 55 degree for the 
highest energy transfer. There is a strong destructive interference 
effect from the Born terms which even leads to a sign change near 
threshold, but their importance becomes smaller and smaller when 
approaching the $\Delta$ region. Its size is only a factor three smaller 
than $f_T$ at the two highest $\omega$ values. 

\begin{figure}
\centerline{%
\epsfysize=18cm
\epsffile{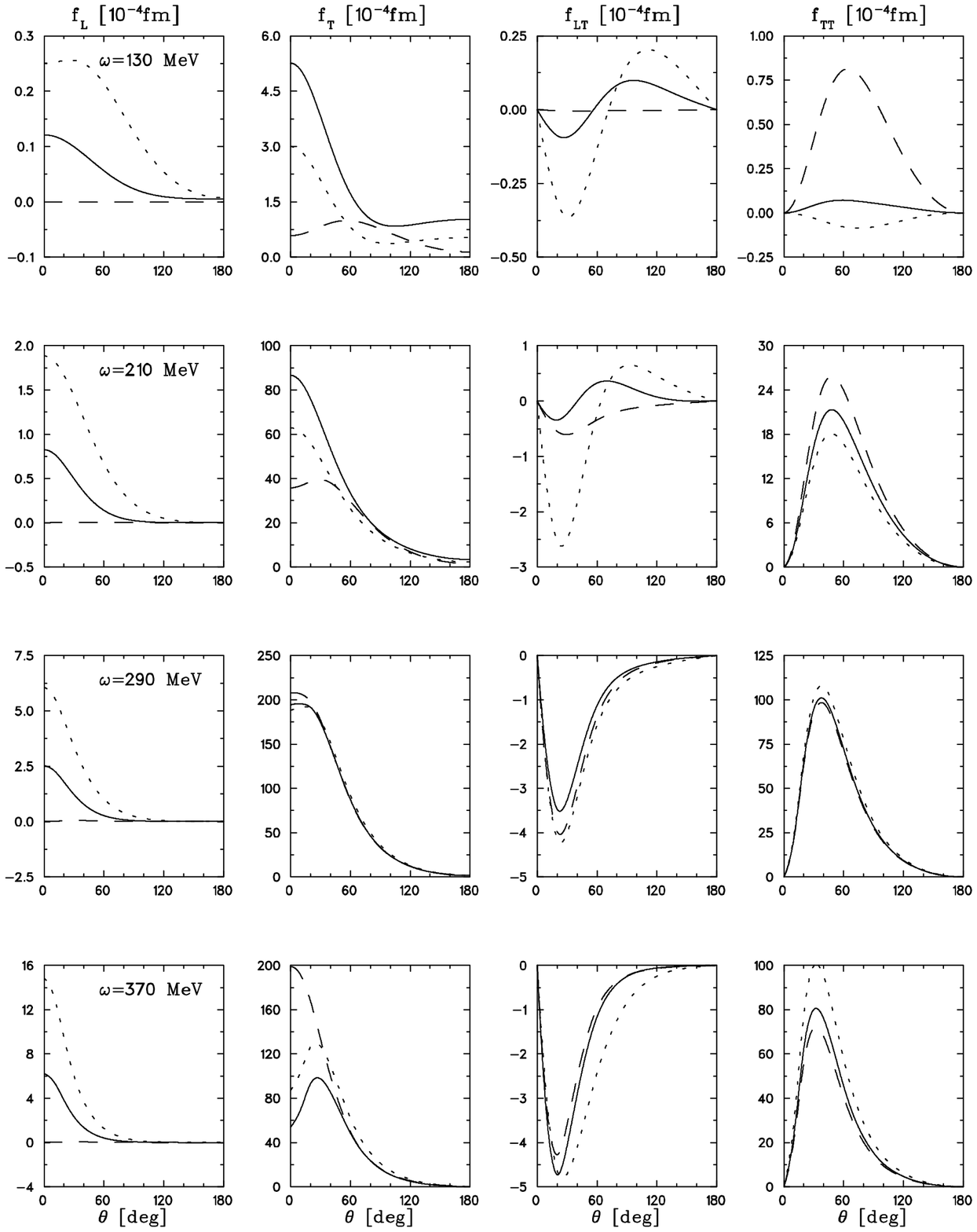}}
\vspace*{.5cm}
\caption{Structure functions for the set B at constant momentum transfer
$k^2=5\,\rm{fm}^{-2}$ for various energy transfers. Notation as in 
Fig.\ \ref{2_fm-2}.}
\label{5_fm-2}
\end{figure}

\begin{figure}
\centerline{%
\epsfysize=18cm
\epsffile{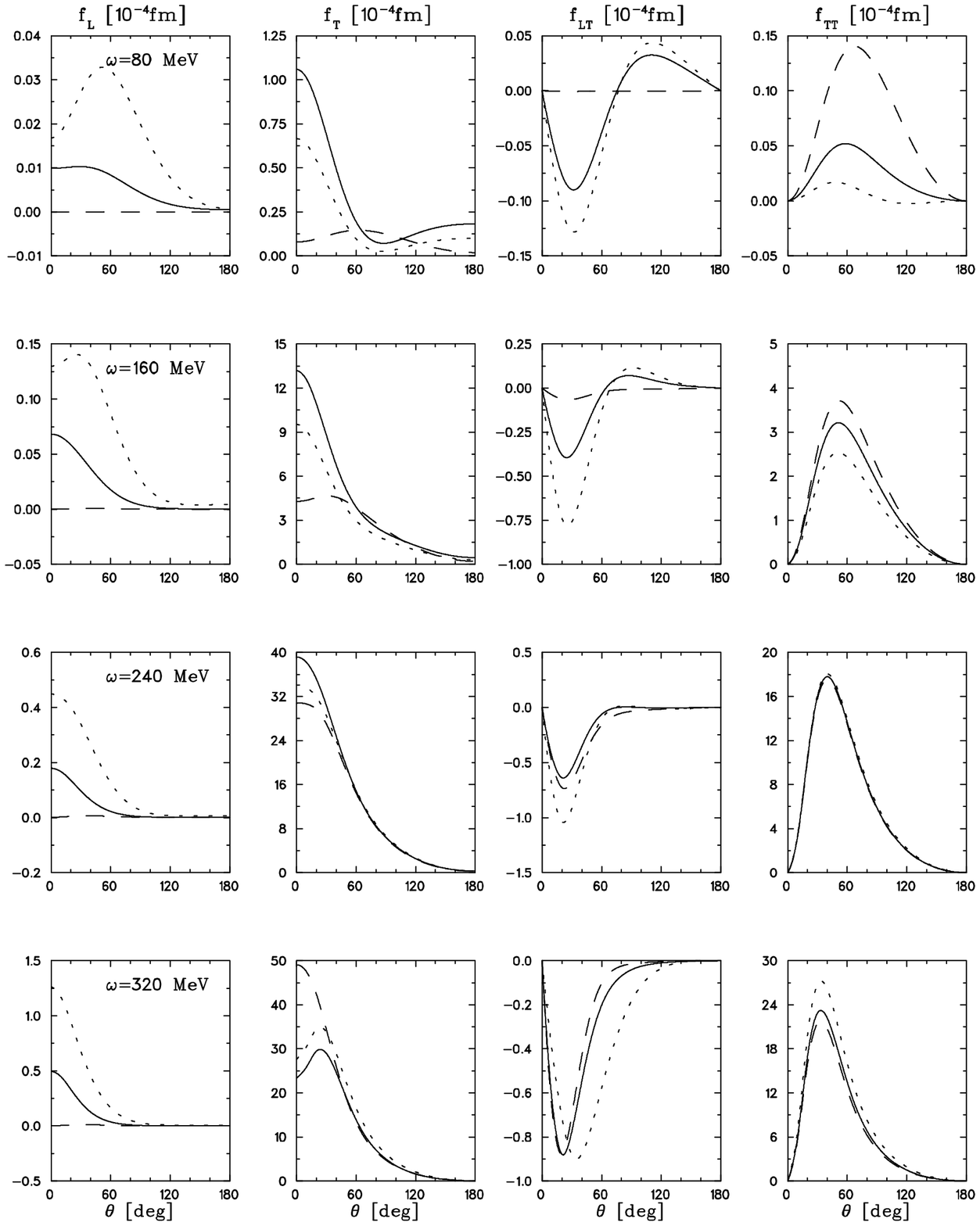}}
\vspace*{.5cm}
\caption{Structure functions for the set C at constant momentum transfer
$k^2=10\,\rm{fm}^{-2}$ for various energy transfers. Notation as in 
Fig.\ \ref{2_fm-2}.}
\label{10_fm-2}
\end{figure}

For the sets B and C at constant three momentum transfers of 
$k^2=5\,\rm{fm}^{-2}$ and $k^2=10\,\rm{fm}^{-2}$, respectively, 
we have chosen the four energy transfers $\omega$ 
in such a manner that pairwise they correspond to the same invariant 
energies $W_{\gamma^\ast d}=$ 2057, 2137, 2217, and 2297 MeV, 
i.e., to the same 
pion momentum. The structure functions are shown in Figs.\ \ref{5_fm-2} 
and \ref{10_fm-2}, respectively. Comparing the corresponding structure 
functions in both figures, one notices a qualitative similarity in the 
shape although they differ substantially in absolute size due 
to the considerably different momentum transfer. Thus we can 
limit the discussion to set B in Fig.\ \ref{5_fm-2}. As above, $f_L$ 
is dominated by the Born terms showing a sizable destructive 
interference between direct and crossed contributions. With increasing 
$\omega$ the absolute size increases and the forward peaking becomes 
more and more pronounced. 

In $f_T$ one clearly notices the increasing importance of the 
$\Delta$ contribution as is expected when crossing the $\Delta$ region. 
The panels at the two lowest energy transfers at $\omega=130$ and 210 
MeV show qualitatively the same behaviour as the panels for $f_T$ 
at $\omega=160$ and 200 MeV of set A in Fig.\ \ref{2_fm-2}. 
Right on the $\Delta$ at 
$\omega=290$ MeV, they have almost no influence. Only above
this region the Born terms become significant again and interfere 
destructively in the forward direction so that the forward peak moves to 
about 30 degree.

With respect to the interference structure functions, $f_{LT}$ shows 
the same qualitative angular behaviour as for the set A for energy 
transfers below the $\Delta$ resonance region. However, on and above 
this region this is not true any more. There the $\Delta$ 
dominates and the Born terms contribute very little. The transverse 
interference structure function $f_{TT}$ exhibits a different behaviour 
compared to set A at all energies. At the lowest energy of 130 MeV, 
one notices an almost complete cancellation between resonance and 
Born contributions. But already at 210 MeV the Born terms become 
relatively small, and at $\omega=290$ MeV they  
are completely negligible while at higher energy transfers they start to 
show again some influence, although direct and crossed Born terms 
interfere destructively, but lead to a slight enhancement of the 
resonance contribution. 

Altogether, all structure functions of sets B and C 
except $f_{L}$ are dominated 
by the $\Delta$ over a large region of energy transfers crossing 
the resonance position. This was to be expected because of the 
dominance of the transverse $\Delta$ excitation current. Only 
$f_{L}$ is governed by the Born terms, a fact which might be changed 
somewhat if the neglected small $C2$ contribution is included. 
But we do not expect drastic changes due to the smallness of such 
contributions. As to the absolute size, $f_{T}$ is by far the largest 
below and on the resonance, followed by $f_{TT}$ which becomes of 
the same size above the resonance. Sizeably smaller are $f_{L}$ and 
$f_{LT}$. 

\begin{figure}[b]
\centerline{%
\epsfysize=5.3cm
\epsffile{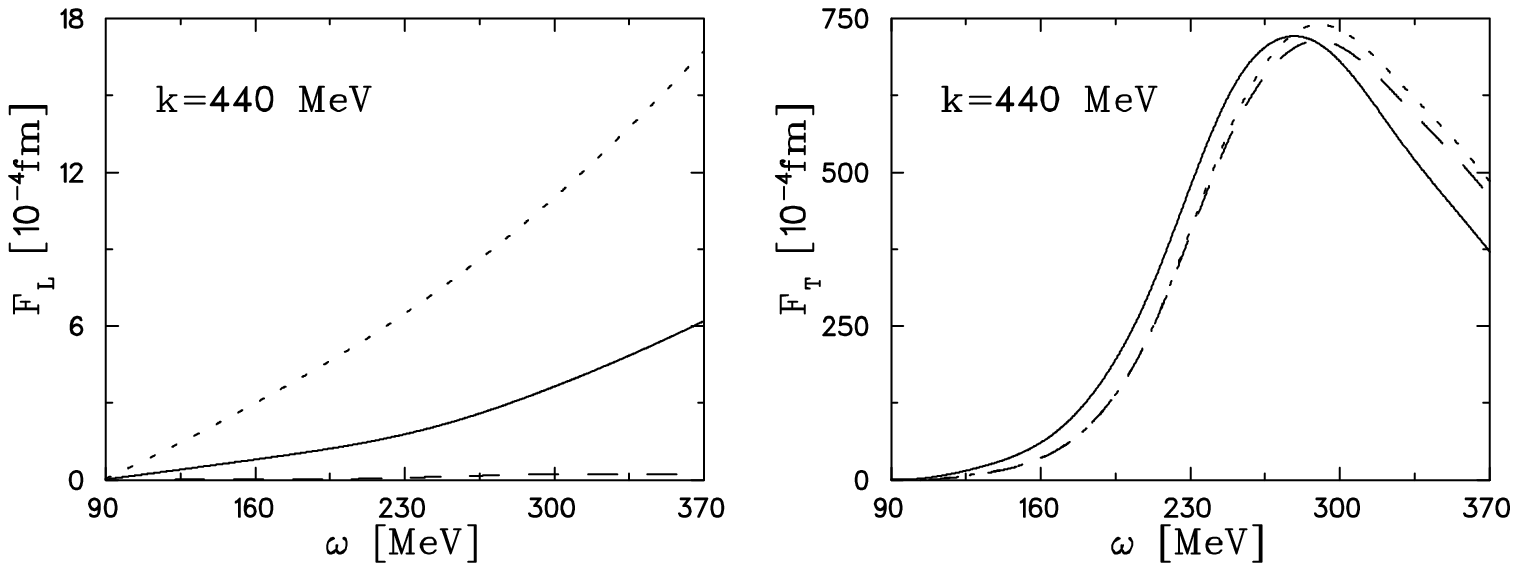}}
\vspace*{.5cm}
\caption{Longitudinal and transverse form factors for set B. Notation as in 
Fig.\ \ref{2_fm-2}.}
\label{Form_5}
\end{figure}

\begin{figure}[t]
\centerline{%
\epsfysize=5.3cm
\epsffile{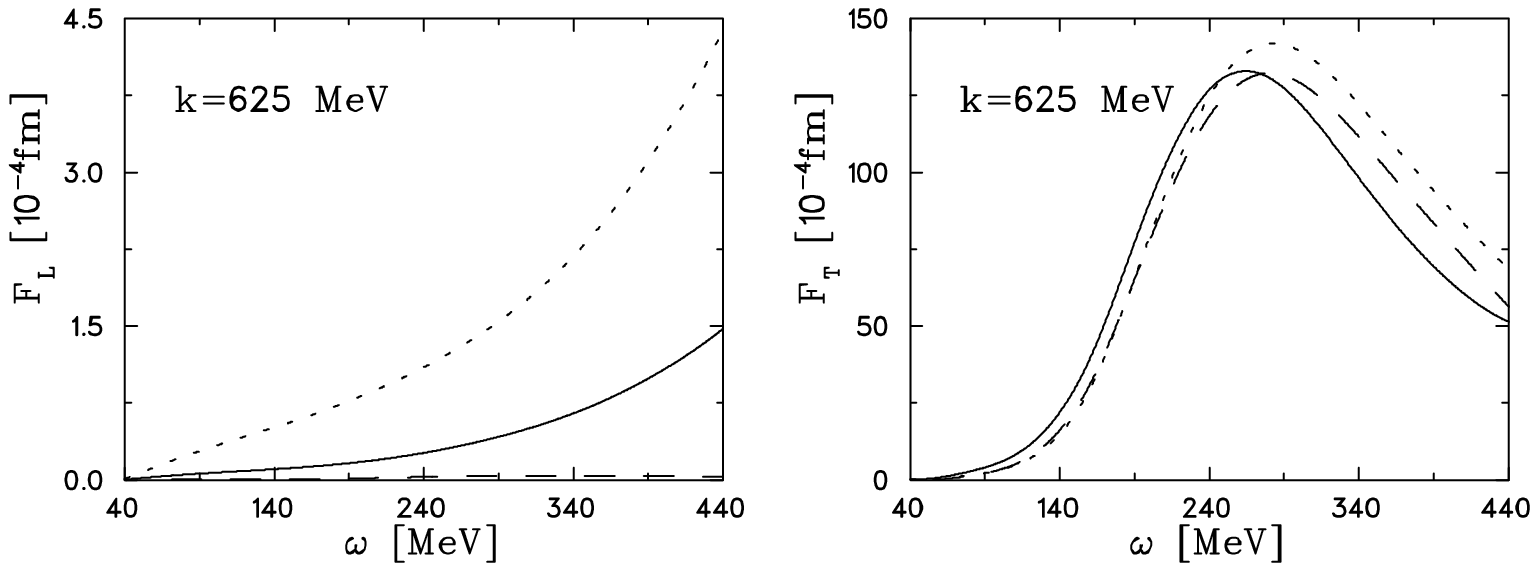}}
\vspace*{.5cm}
\caption{Longitudinal and transverse form factors for set C. Notation as in 
Fig.\ \ref{2_fm-2}.}
\label{Form_10}
\end{figure}

We also have calculated the form factors $F_L$ and $F_T$ of 
the inclusive cross section for the sets B and C, because they allow 
a better comparison of the absolute values. They are shown in Figs.\
\ref{Form_5} and \ref{Form_10}. For both sets the longitudinal form 
factor $F_L$ shows a steady increase over the whole range of energy 
transfers considered which arises essentially from the e.m.\ form 
factor because one approaches the photon point with increasing $\omega$.
By destructive interference the direct Born contribution is reduced 
to about 30 percent via the crossed one. As noted before, the 
$\Delta$ is negligible here in contrast to the transverse form factor 
$F_T$. Here the direct and crossed Born contribution show a distinctly 
different behaviour. While the direct one leads only to a slight general 
increase of the $\Delta$ contribution, the crossed Born contribution 
results in a downshift of the maximum by about 20 MeV and a significant 
reduction above the maximum. In absolute size, $F_T$ is about one to 
two orders of magnitude bigger than $F_L$. Comparing sets B and C one 
notices a decrease of the form factors by a factor of about 5 due to 
the higher four momentum transfers involved in set C. 

\begin{figure}
\centerline{%
\epsfysize=18cm
\epsffile{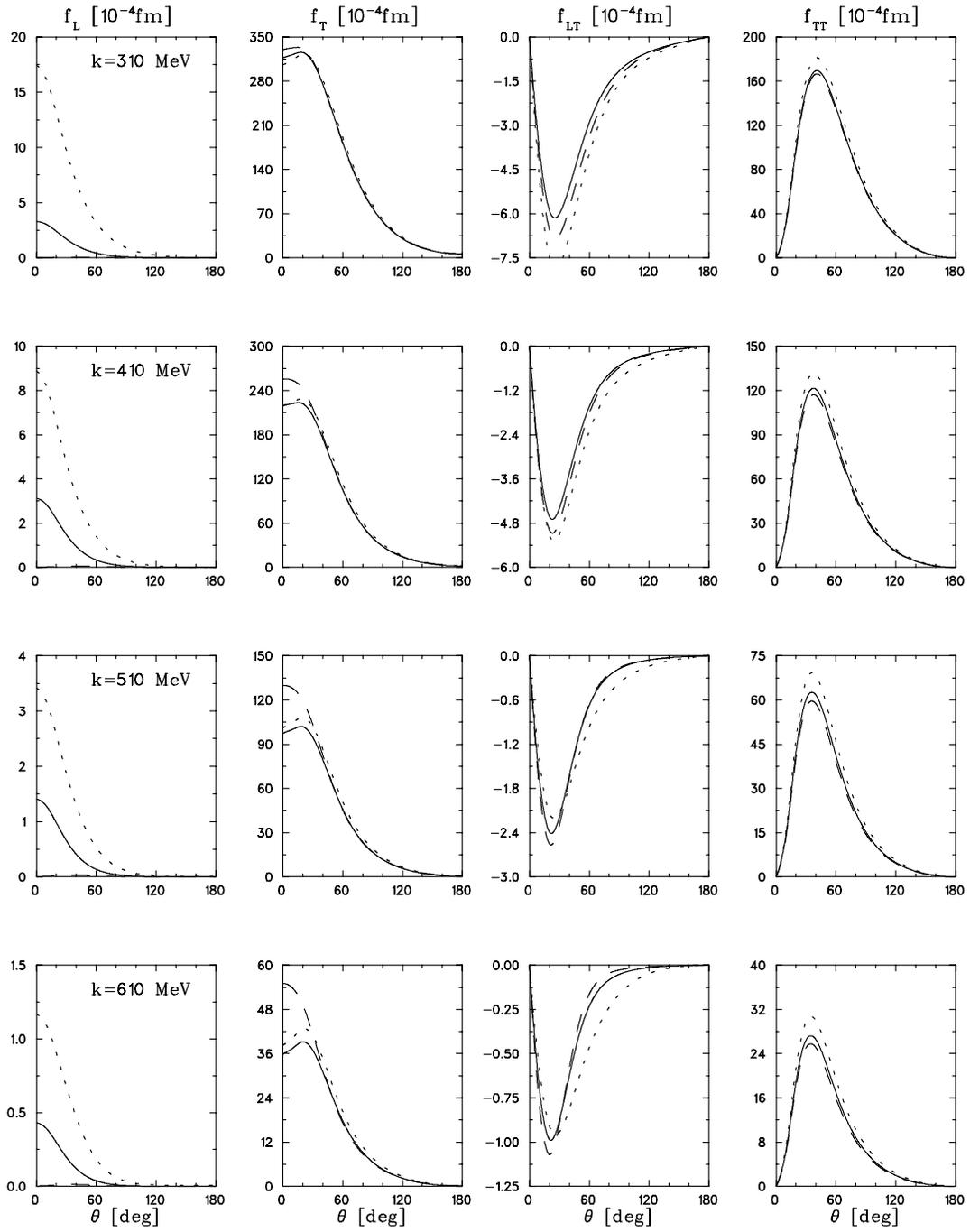}}
\vspace*{.5cm}
\caption{Structure functions for the set D at constant energy transfer
$\omega=300$ MeV for various momentum transfers. Notation as in 
Fig.\ \ref{2_fm-2}.}
\label{300_MeV}
\end{figure}

\begin{figure}
\centerline{%
\epsfysize=5.3cm
\epsffile{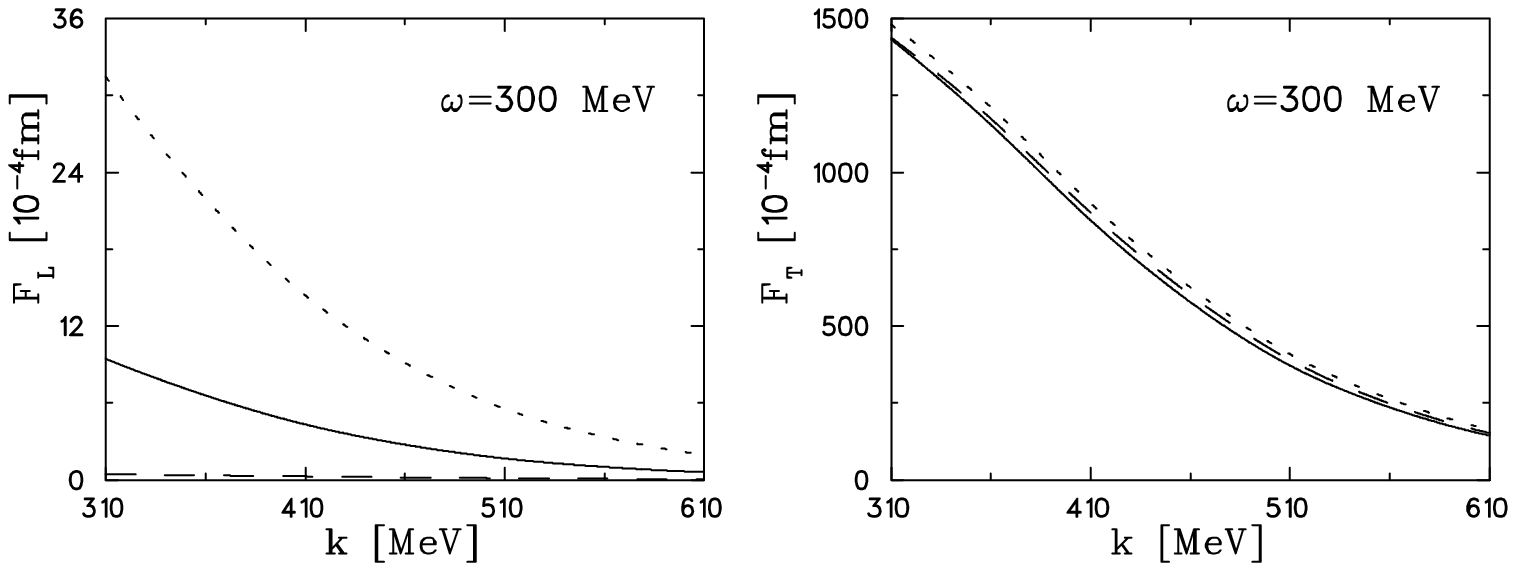}}
\vspace*{.5cm}
\caption{Longitudinal and transverse form factors for set D. Notation as in 
Fig.\ \ref{2_fm-2}.}
\label{Form_300}
\end{figure}

Finally, we will consider in Fig.\ \ref{300_MeV} the structure functions 
of set D at a constant energy transfer 
$\omega=300$ MeV with varying three momentum transfers, which 
means close and slightly above the resonance region. Accordingly, 
one readily notices the angular behaviour which we had found before 
at corresponding kinematics. In $f_{L}$ the destructive interference 
of direct and crossed Born terms becomes again apparent. For $f_{T}$ 
one readily sees the increasing reduction of the forward $\Delta$ peak 
by the background contributions with increasing momentum transfer 
resulting in a small shift away from the forward direction. 
On the contrary, the 
interference structure functions show qualitatively very little 
variation with the momentum transfer. Only the absolute size 
decreases rapidly which is also the case for the diagonal structure 
functions $f_{L}$ and $f_{T}$ and for the corresponding inclusive 
form factors shown in Fig.\ \ref{Form_300}. This is a consequence of the 
increasing four momentum transfer which governs the e.m.\ form factors 
of the elementary vertices.

\section{Summary and Conclusions}
Coherent $\pi^0$ electroproduction on the deuteron has been studied in 
the impulse approximation neglecting rescattering and two-body effects. 
For the elementary reaction on the nucleon, we have considered besides the 
dominant excitation of the $\Delta$(1232) resonance the usual Born terms, 
essentially in the same frame work as in the analogous photo process. 
The four structure functions which govern the differential cross section
have been evaluated along various cuts in the plane of energy-momentum 
transfer. In particular, the relative importance of the resonance and 
background contributions have been studied in detail. They show quite 
different influences in the different structure functions, due to the 
fact, that the excitation of the $\Delta$ resonance proceeds dominantly via 
M1 excitations, which are purely transverse. It now remains as a task for 
the future to study the influence of rescattering and, furthermore, to 
use a more refined elementary production operator, in particular for the 
threshold region.

\renewcommand{\theequation}{A.\arabic{equation}}
\setcounter{equation}{0}
\section*{Appendix A: Completeness proof for the basic operator set}
In this appendix we will present an independent proof that the operators 
listed in Table \ref{tabkqls} form a complete and independent basis for 
the representation of the $T$-matrix of coherent e.m.\ pseudoscalar 
production on a spin-one target. Independence means that a relation
\beq
\sum_\alpha f_\alpha(k,q,\cos \theta)\,{\cal O}_\alpha\equiv 0 
\eeq
with $k=|\vec k|$, $q=|\vec q\,|$ and $\cos\theta=\hat k\cdot \hat q$, is 
fulfilled only if $f_\alpha\equiv 0$ for all $\alpha$. 

To this end  we consider first an alternative set whose elements also 
have to be built out of the only available unit vectors $\hat k$ and 
$\hat q$ and the spin-one operators $S^{[\Sigma]}$ $(\Sigma=0,1,2)$.
Explicitly these new operators have the form 
$\vec O_{(KQ)L}^{\ \Sigma}\cdot\vec\epsilon_{\mu}$ with
\begin{equation}
\vec O_{(KQ)L}^{\ \Sigma}:=[A_{KQ}^{[L]}\times S^{[\Sigma]}]^{[1]}\,,
\label{sigmakql}
\end{equation}
where we have introduced 
\beq
A_{KQ}^{[L]}:=[ k^{[K]}\times q^{[Q]}]^{[L]}\,\label{AKQL}
\eeq
with the following recursive definitions of the spherical tensors $a^{[n]}$
\begin{eqnarray}
a^{[0]} &:=& 1,\nonumber\\
a^{[1]} &:=& \hat a,\nonumber\\
a^{[n+1]} &:=& [ a^{[1]}\times a^{[n]}]^{[n+1]}\,,\quad\forall\ n\in
\mathbf{N}\nonumber
\,.
\end{eqnarray} 
They are related to the spherical harmonics by
\beq
a^{[n]}=\alpha_n\,Y^{[n]}\quad \mbox{ with }\quad 
\alpha_n=\sqrt{\frac{4\pi\,n!}{(2n+1)!!}}\,.
\eeq
Angular momentum coupling rules restrict the $L$-values in (\ref{sigmakql}) 
to $|\Sigma-1|\leq L\leq \Sigma+1$. 
The possible combinations are listed in Table \ref{SigmaL}.
Furthermore, in order to insure that $\vec O_{(KQ)L}^{\ \Sigma}$ be a 
pseudo-vector, the condition $K+Q=\,$even has to be fulfilled. Taking into 
account this condition and the coupling rules, one finds for the values of the 
indices $K,Q,L$ the combinations listed in Table \ref{KQL}.

As next we will derive a recursion relation for the $A_{KQ}^{[L]}$ 
based on the recoupling of the following expression
\beqa
A_{NN}^{[0]}\,A_{KQ}^{[L]}&=&[ A_{NN}^{[0]}\times A_{KQ}^{[L]}]^{[L]}
\nonumber\\
&=&\sum_{K_{1}=|K-N|}^{K+N}\,\sum_{Q_{1}=|Q-N|}^{Q+N}C_{NKQ}^{(K_{1}Q_{1})L}
A_{K_{1}Q_{1}}^{[L]}\,,\label{recursion}
\eeqa
where
\begin{eqnarray}
C_{NKQ}^{(K_{1}Q_{1})L}&=&
(-)^{N+L+K_{1}+Q}\,\frac{\hat K_{1}\hat Q_{1}}{\hat N}\,
\beta_{NK}^{K_{1}}\,\beta_{NQ}^{Q_{1}}\,
\left\{
\begin{array}{ccc}
K     & Q    & L \\
Q_{1} & K_{1}& N
\end{array}
\right\}\,,\nonumber\\
\beta_{AB}^{C}&=& (-)^C\sqrt{\frac{A!\,B!\,(2C+1)!!}{(2A+1)!!\,(2B+1)!!\,C!}}
\left(
\matrix{
A     & B    & C \cr
0     & 0    & 0 \cr
}
\right)
\,.
\end{eqnarray}
One should note  that $A_{NN}^{[0]}$ is basically the Legendre polynomial 
$P_{N}(\cos\theta)$, namely
\begin{equation}
A_{NN}^{[0]}=\alpha_{N}^{2}\frac{(-)^{N}\hat N}{4\pi}P_{N}(\cos\theta)\,.
\end{equation}
Separating in (\ref{recursion}) on the rhs the term with the highest indices 
$K_1=K+N$ and $Q_1=Q+N$, one finds 
\begin{eqnarray}
A_{NN}^{[0]}\, A_{KQ}^{[L]}
&=&\sum_{K_{1}=|K-N|}^{K+N-1}\,\sum_{Q_{1}=|Q-N|}^{Q+N}
C_{NKQ}^{(K_{1}Q_{1})L}\,A_{K_{1}Q_{1}}^{[L]}\,\nonumber\\
&&+\sum_{Q_{1}=|Q-N|}^{Q+N-1}C_{NKQ}^{((K+N)Q_{1})L}\,A_{(K+N)Q_{1}}^{[L]}\,
\nonumber\\
&&+C_{NKQ}^{((K+N)(Q+N))L}\,A_{(K+N)(Q+N)}^{[L]}\,.
\end{eqnarray}
With the replacements $K+N\rightarrow K$ and $Q+N\rightarrow Q$ 
the above equation finally leads to the recursion relation
\begin{eqnarray}
C_{N(K-N)(Q-N)}^{(KQ)L}A_{KQ}^{[L]}
&=&A_{NN}^{[0]}A_{(K-N)(Q-N)}^{[L]}\nonumber\\
&&-\sum_{K_{1}=|K-2N|}^{K-1}\sum_{Q_{1}=|Q-2N|}^{Q}
C_{N(K-N)(Q-N)}^{(K_{1}Q_{1})L}\,A_{K_{1}Q_{1}}^{[L]}\,\nonumber\\
&&-\sum_{Q_{1}=|Q-2N|}^{Q-1}C_{N(K-N)(Q-N)}^{(KQ_{1})L}\,A_{KQ_{1}}^{[L]}\,.
\label{recursiona}
\end{eqnarray}
Since $K,Q$ and $L$ have to fulfil the usual triangular relation N is 
limited by $N\le\frac{1}{2}(K+Q-L)$. 
By this relation, $A_{KQ}^{[L]}$ is expressed as a linear superposition of 
$A_{K_1Q_1}^{[L]}$ where $K_1\leq K$, $Q_1\leq Q$, and $K_1+Q_1<K+Q$. 

According to Table \ref{KQL} one has for $L\in\{0,1\}$ only 
the combination $K=Q$. While for $L=0$ the recursion formula becomes trivial 
\begin{equation}
A_{NN}^{[0]}=A_{NN}^{[0]}A_{00}^{[0]}\,,
\end{equation} 
one finds for $L=1$
\begin{eqnarray}
C_{N(K-N)(K-N)}^{(KK)1}A_{KK}^{[1]}
&=&A_{NN}^{[0]}A_{(K-N)(K-N)}^{[1]}\nonumber\\
&-&\sum_{K_{1}=|K-2N|}^{K-1}\sum_{Q_{1}=|K-2N|}^{K}C_{N(K-N)(K-N)}^{(K_{1}Q_{1})1}A_{K_{1}Q_{1}}^{[1]}\,\nonumber\\
&-&\sum_{Q_{1}=|K-2N|}^{K-1}C_{N(K-N)(K-N)}^{(KQ_{1})1}A_{KQ_{1}}^{[1]}\,.
\label{recursion1}
\end{eqnarray}
Exploiting the properties of the coefficients $C_{NKQ}^{(K_{1}Q_{1})L}$ 
($N+K+K_{1}$ and $N+Q+Q_{1}$ must both be even numbers while $K_{1},Q_{1}$ 
and $1$ must obey the triangular relation) (\ref{recursion1}) simplifies to 
\begin{equation}
C_{N(K-N)(K-N)}^{(KK)1}A_{KK}^{[1]}=A_{NN}^{[0]}A_{(K-N)(K-N)}^{[1]}-\sum_{K_{1}=|K-2N|}^{K-2}C_{N(K-N)(K-N)}^{(K_{1}K_{1})1}A_{K_{1}K_{1}}^{[1]}\,.
\end{equation}
Setting $N=1$, one finds
\begin{equation}
C_{1(K-1)(K-1)}^{(KK)1}A_{KK}^{[1]}=A_{11}^{[0]}A_{(K-1)(K-1)}^{[1]}
-C_{1(K-1)(K-1)}^{((K-2)(K-2))1}A_{(K-2)(K-2)}^{[1]}\,,
\end{equation}
from which readily follows that $A_{11}^{[1]}$ is the only initial element 
of the recursion for $L=1$ as was $A_{00}^{[0]}$ for $L=0$. Thus every 
$A_{KK}^{[1]}$ is related to $A_{11}^{[1]}$ times a function in $\cos\theta$.

Proceeding in complete analogy, we obtain for $L=2$ and $N=1$
\begin{eqnarray}
C_{1(K-1)(Q-1)}^{(KQ)2}\,A_{KQ}^{[2]}
&=&A_{11}^{[0]}\,A_{(K-1)(Q-1)}^{[2]}-\sum_{K_{1}=|K-2|}^{K-1}
\sum_{Q_{1}=|Q-2|}^{Q-1}C_{1(K-1)(Q-1)}^{(K_{1}Q_{1})2}\,A_{K_{1}Q_{1}}^{[2]}
\,\nonumber\\
&&-\sum_{K_{1}=|K-2|}^{K-1}C_{1(K-1)(Q-1)}^{(K_{1}Q)2}\,A_{K_{1}Q}^{[2]}
-\sum_{Q_{1}=|Q-2|}^{Q-1}C_{1(K-1)(Q-1)}^{(KQ_{1})2}\,A_{KQ_{1}}^{[2]}\,,
\end{eqnarray}
respectively,
\begin{eqnarray}
C_{1(K-1)(Q-1)}^{(KQ)2}A_{KQ}^{[2]}
&=&A_{11}^{[0]}A_{(K-1)(Q-1)}^{[2]}-C_{1(K-1)(Q-1)}^{((K-2)(Q-2))2}\,
A_{(K-2)(Q-2)}^{[2]}\,\nonumber\\
&&-C_{1(K-1)(Q-1)}^{((K-2)Q)2}\,A_{(K-2)Q}^{[2]}
-C_{1(K-1)(Q-1)}^{(K(Q-2))2}\,A_{K(Q-2)}^{[2]}\,.\label{recurs2}
\end{eqnarray}
As independent elements one finds here 
$A_{11}^{[2]},A_{20}^{[2]},A_{02}^{[2]}$ which cannot be further reduced. 
In other words, every $A_{KQ}^{[2]}$ can be expressed as a linear 
combination of these independent elements with appropriate functions 
in $\cos\theta$ as coefficients. An analogous relation holds for $L=3$ 
\begin{eqnarray}
C_{1(K-1)(Q-1)}^{(KQ)3}A_{KQ}^{[3]}
&=&A_{11}^{[0]}A_{(K-1)(Q-1)}^{[3]}-C_{1(K-1)(Q-1)}^{((K-2)(Q-2))3}\,
A_{(K-2)(Q-2)}^{[3]}\,\nonumber\\
&&-C_{1(K-1)(Q-1)}^{((K-2)Q)3}\,A_{(K-2)Q}^{[3]}
-C_{1(K-1)(Q-1)}^{(K(Q-2))3}\,A_{K(Q-2)}^{[3]}\,.
\end{eqnarray}
Here the remaining independent elements are 
$A_{22}^{[3]},A_{31}^{[3]},A_{13}^{[3]}$. With this we finally have proven
that the set of operators listed in Table \ref{Deuteron} is independent 
and complete. 

Although these basic elements do not separate in general into transverse 
and longitudinal operators this set provides a better starting point for 
a systematic (computer aided) decomposition of the amplitude. The transition
to the transverse and longitudinal operators of Table \ref{tabkqls} can then 
be obtained by the following transformation
\begin{eqnarray}
{\cal O}_{T,1}
&=&
-i\sqrt{2}\ {\cal O}_{1}
\,,\nonumber\\ 
({\cal O}_{T,2},  {\cal O}_{T,3},  {\cal O}_{T,4},  {\cal O}_{L,1},  
{\cal O}_{L,2} )^{T}
&=&{\cal A} \cdot ({\cal O}_{2}, {\cal O}_{3},  {\cal O}_{4},  
{\cal O}_{5},  {\cal O}_{6} )^{T}\,,\nonumber\\ 
({\cal O}_{T,5}, {\cal O}_{T,6},  {\cal O}_{T,7},  {\cal O}_{T,8},  
{\cal O}_{T,9}, {\cal O}_{L,3},  {\cal O}_{L,4} )^{T}
&=&
{\cal B} \cdot ( {\cal O}_{7}, {\cal O}_{8}, {\cal O}_{9}, {\cal O}_{10}, 
{\cal O}_{11}, {\cal O}_{12}, {\cal O}_{13})^{T}\,.
\end{eqnarray}
with the transformation matrices
\begin{eqnarray}
\cal{A}&=&
\left(
\begin{array}{ccccc}
\frac{2(1-P_{2}(\cos\theta ))}{9} & 0 & -\frac{2\sqrt{5}}{\sqrt{3}}
\cos\theta  & \sqrt{\frac{5}{3}} & \sqrt{\frac{5}{3}} \\
-\frac{2}{3} & 0 & 0 & 0 & -\sqrt{\frac{5}{3}} \\
-\frac{2}{3}\cos\theta  & -1 & -\sqrt{\frac{5}{3}} & 0 & 0 \\
\frac{1}{3} & 0 & 0 & 0 &  -\sqrt{\frac{5}{3}} \\
\frac{1}{3}\cos\theta  & 1 & -\sqrt{\frac{5}{3}} & 0 & 0 
\end{array}
\right)\,,
\\
\cal{B}&=&
\left(
\begin{array}{ccccccc}
-\frac{i\sqrt{2}}{\sqrt{15}} & -\frac{i\sqrt{10}}{3} & 0 & 
\frac{i\sqrt{10}}{3}\cos\theta  & 0 & 0 & -\frac{i\sqrt{28}}{3} \\
-\frac{i\sqrt{2}}{\sqrt{15}}\cos\theta  & 0 & -\frac{i\sqrt{5}}{3\sqrt{2}} & 
\frac{i\sqrt{5}}{3\sqrt{2}} & -\frac{i\sqrt{7}}{\sqrt{3}} & 0 & 0 \\
-\frac{i\sqrt{2}}{\sqrt{15}} & \frac{i\sqrt{10}}{3} & 
\frac{i\sqrt{10}}{3}\cos\theta  & 0 & 0 & -\frac{i\sqrt{28}}{3} & 0 \\
0 & 0 & 0 & -\frac{i\sqrt{3}}{\sqrt{2}} & 0 & 0 & 0 \\
\frac{i}{\sqrt{2}} & -\frac{i3\sqrt{5}}{\sqrt{2}} & 0 & 0 & 0 & 0 & 0 \\
-\frac{i3}{2\sqrt{2}} & -\frac{i(3\sqrt{15}-4)}{\sqrt{360}} & 0 & 
-\frac{i\sqrt{5}}{3\sqrt{2}}\cos\theta  & 0 & 0 & -\frac{i\sqrt{28}}{3} \\
\frac{i\sqrt{3}}{\sqrt{10}}\cos\theta  & 
-\frac{i\sqrt{5}}{\sqrt{2}}\cos\theta  & 
-\frac{i\sqrt{5}(\sqrt{3}-1)}{3\sqrt{2}} & 
\frac{i\sqrt{5}}{3\sqrt{2}} & -\frac{i\sqrt{7}}{\sqrt{3}} & 0 & 0
\end{array}
\right)\,.
\end{eqnarray}

\begin{table}[h]
\caption{Basic set of independent, nonrelativistic transverse and 
longitudinal operators, where $\hat k =\vec k/|\vec k|$ and 
$\hat q =\vec q/|\vec q\,|$.}
\begin{center}
\renewcommand{\arraystretch}{1.4}
\begin{tabular}{cll}
 $\beta$ & ${\cal O}_{T,\beta} $ & ${\cal O}_{L,\beta}$ \\  
\hline
1 & $\vec \epsilon\cdot (\hat k \times \hat q\,) $ & 
$\hat k \cdot \vec S $\\
2 & $\vec \epsilon\cdot (\hat k \times \hat q\,)\,
(\hat k \times \hat q\,)\cdot \vec S  $ &
$\hat q \cdot \vec S$ \\
3 & $\vec \epsilon\cdot (\hat k \times (\hat k \times \vec S\,))$ & 
$[(\hat k\times \hat q\,)\times \hat k\,]^{[2]}\cdot S^{[2]}  $ \\
4 & $\vec \epsilon\cdot (\hat k \times (\hat q \times \vec S\,))$ &
$[(\hat k\times \hat q\,)\times \hat q\,]^{[2]}\cdot S^{[2]} $\\
5 & $\vec \epsilon\cdot (\hat k \times \hat q\,)\,\hat k^{[2]}\cdot S^{[2]}$ 
& \\
6 & $\vec \epsilon\cdot (\hat k \times \hat q\,)\,[\hat k\times \hat q]^{[2]}
\cdot S^{[2]}  $ & \\
7 & $\vec \epsilon\cdot (\hat k \times \hat q\,)\, \hat q^{[2]}\cdot S^{[2]} 
 $ & \\
8 & $\vec \epsilon\cdot (\hat k \times [\hat k\times S^{[2]}]^{[1]})$ & \\
9 & $\vec \epsilon\cdot (\hat k \times [\hat q\times S^{[2]}]^{[1]})$ & \\
\end{tabular}
\end{center}
\label{tabkqls}
\end{table}

\begin{table}[h]
\caption{Possible values of $L$ and $\Sigma$ of the operators in (\ref{AKQL}).}
\begin{center}
\begin{tabular}{ccccccccccc}
$\Sigma$ &  & 0 &  &\multicolumn{3}{c}{1}&  &\multicolumn{3}{c}{2}\\
\hline
$L$      &  & 1 &  & 0 & 1 & 2 &  & 1 & 2 & 3\\
\end{tabular}
\end{center}
\label{SigmaL}
\end{table}

\begin{table}[h]
\caption{Listing of possible $(KQL)$ combinations.}
\begin{center}
\begin{tabular}{ccc}
$L$      &  $K$  &  $Q$      \\
\hline
 0,1            & $K\ge 0$     & $K$    \\ 
\hline
 2              & 0            & 2      \\
                & 1            & 1,3     \\
                & $K\ge 2$     & $K-2,\,K,\,K+2$    \\
\hline
 3              & $K<3$        & $K+2$   \\
                & $K\ge 3$     & $K-2,\,K,\,K+2$    \\
\end{tabular}
\end{center}
\label{KQL}
\end{table}

\begin{table}[h]
\caption{Listing of the 13 basic operators 
${\cal O}_{\alpha}=\vec O_{(KQ)L}^{\ \Sigma}\cdot\vec\epsilon_{\mu}$.}
\begin{center}
\begin{tabular}{cccccccccccccc}
$\alpha$       & 1 & 2 & 3 & 4 & 5 & 6 & 7 & 8 & 9 & 10 & 11 & 12 & 13 \\
\hline
$\Sigma$  & 0 & 1 & 1 & 1 & 1 & 1 & 2 & 2 & 2 & 2 & 2 & 2 & 2 \\
\hline
$L$      & 1 & 0 & 1 & 2 & 2 & 2 & 1 & 2 & 2 & 2 & 3 & 3 & 3 \\
$K$      & 1 & 0 & 1 & 1 & 0 & 2 & 1 & 1 & 0 & 2 & 2 & 1 & 3 \\
$Q$      & 1 & 0 & 1 & 1 & 2 & 0 & 1 & 1 & 2 & 0 & 2 & 3 & 1 \\
\end{tabular}
\end{center}
\label{Deuteron}
\end{table}

\end{document}